\documentclass[sigconf, dvipsnames]{acmart}
\pdfoutput=1
\usepackage{geometry}
\usepackage{graphicx}
\usepackage{pgfplots}
\pgfplotsset{compat=1.17}
\usepackage{tikz-cd}
\usepackage{tikz}
\usepackage{hyperref}
\usepackage{lipsum}
\usepackage{listings}
\usepackage{subcaption}
\usepackage{geometry}
\usepackage{enumitem}
\usepackage{xcolor}
\usepackage{caption}

\usepackage{pgfplotstable}
\usepgfplotslibrary{colorbrewer}
\pgfplotsset{compat=1.17}

\usepgfplotslibrary{groupplots}
\let\origthelstnumber\thelstnumber
\makeatletter
\newcommand*\Suppressnumber{%
  \lst@AddToHook{OnNewLine}{%
    \let\thelstnumber\relax%
     \advance\c@lstnumber-\@ne\relax%
    }%
}

\newcommand*\Reactivatenumber[1]{%
  \lst@AddToHook{OnNewLine}{%
   \let\thelstnumber\origthelstnumber%
   \setcounter{lstnumber}{\numexpr#1-1\relax}%
   }%
}

\pgfplotsset{compat=1.17}
 \usepackage{etoolbox}
\makeatletter
\patchcmd{\@verbatim}
  {\verbatim@font}
  {\verbatim@font\small}
  {}{}
\makeatother

\hypersetup{
   breaklinks=true,   % splits links across lines
   colorlinks=true,   % displays links as colored text instead of blocks
}

\AtBeginDocument{%
  \providecommand\BibTeX{{%
    \normalfont B\kern-0.5em{\scshape i\kern-0.25em b}\kern-0.8em\TeX}}}

\setcopyright{acmcopyright}
\copyrightyear{2023}

\acmYear{2024}\copyrightyear{2024}
\acmConference[AI4Sys '24]{Workshop on AI For Systems}{June 3--7, 2024}{Pisa, Italy}
\acmBooktitle{Workshop on AI For Systems (AI4Sys '24), June 3--7, 2024, Pisa, Italy}
\acmDOI{10.1145/3660605.3660944}
\acmISBN{979-8-4007-0652-3/24/06}

\def\BibTeX{{\rm B\kern-.05em{\sc i\kern-.025em b}\kern-.08em
    T\kern-.1667em\lower.7ex\hbox{E}\kern-.125emX}}

\newcommand{\hpcorpus}[0]{\textsc{HPCorpus}}
\newcommand{\mpicodecorpus}[0]{\textsc{MPICodeCorpus}}
\newcommand{\mpicorpus}[0]{\textsc{HPCorpusMPI}}
\newcommand{\mono}[0]{\textsc{MonoCoder}}
\newcommand{\mpir}[0]{\textsc{MPIrigen}}
\newcommand{\tokom}[0]{\textsc{TokomPiler}}

\setcopyright{none}
\renewcommand\footnotetextcopyrightpermission[1]{} % removes footnote with conference information in first column
\begin{document}

\title{\mpir{}: MPI Code Generation through\\Domain-Specific Language Models}

  % Nadav Schneider\textsuperscript{1,2},
  % Niranjan Hasabnis\textsuperscript{3},
  % Vy A. Vo\textsuperscript{3},
  % Tal Kadosh\textsuperscript{1,2},
  % Neva Krien, \\
  % Mihai Capota\textsuperscript{3}, 
  % Abdul Wasay\textsuperscript{3},
  % Guy Tamir\textsuperscript{4},
  % Ted Willke\textsuperscript{3}, 
  % Nesreen Ahmed\textsuperscript{3},
  % Yuval Pinter\textsuperscript{1}, \\
  % Timothy Mattson and 
  % Gal Oren\textsuperscript{5,6}

\author{Nadav Schneider}
\email{nadavsch@post.bgu.ac.il}
\affiliation{%
  \institution{Ben-Gurion University, IAEC}
  \country{Israel}
}

\author{Niranjan Hasabnis}
\email{niranjan.hasabnis@intel.com}
\affiliation{%
  \institution{Intel Labs}
  \country{United States}
}

\author{Vy A. Vo}
\email{vy.vo@intel.com}
\affiliation{%
  \institution{Intel Labs}
  \country{United States}
}

\author{Tal Kadosh}
\email{talkad@post.bgu.ac.il}
\affiliation{%
  \institution{Ben-Gurion University, IAEC}
  \country{Israel}
}

\author{Neva Krien}
\email{nevo.krien@gmail.com}
\affiliation{%
  \institution{Independent Researcher}
  \country{Israel}
}

\author{Mihai Capotă}
\email{mihai.capota@intel.com}
\affiliation{%
  \institution{Intel Labs, United States}
  \country{United States}
}

\author{Guy Tamir}
\email{guy.tamir@intel.com}
\affiliation{%
  \institution{Intel}
  \country{United States}
}

\author{Ted Willke}
\email{ted.willke@intel.com}
\affiliation{%
  \institution{Intel Labs}
  \country{United States}
}

\author{Nesreen Ahmed}
\email{nesreen.k.ahmed@intel.com}
\affiliation{%
  \institution{Intel Labs}
  \country{United States}
}

\author{Yuval Pinter}
\email{pintery@bgu.ac.il}
\affiliation{%
  \institution{Ben-Gurion University}
  \country{Israel}
}

\author{Timothy Mattson}
\email{tim@timmattson.com}
\affiliation{%
  \institution{Independent Researcher}
  \country{United States}
}

\author{Gal Oren}
\email{galoren@cs.technion.ac.il}
\affiliation{%
  \institution{Technion, NRCN}
  \country{Israel}
}

\begin{abstract}
The imperative need to scale computation across numerous nodes highlights the significance of efficient parallel computing, particularly in the realm of Message Passing Interface (MPI) integration. While MPI serves as a cornerstone for large-scale parallelism, its seamless integration into codebases, especially concerning domain decomposition, has proven challenging. Static tools aimed at addressing this challenge have exhibited limited effectiveness and scalability. On the other hand, contemporary language models designed for programming problems have demonstrated utility in parallel programming tasks such as OpenMP pragma generation. However, the challenging parallel programming task of generating MPI-based parallel programs has remained unexplored.

This study first investigates the performance of state-of-the-art language models in generating MPI-based parallel programs.
Findings reveal that widely used models such as GPT-3.5 and PolyCoder (specialized multi-lingual code models) exhibit notable performance degradation when generating MPI-based programs compared to general-purpose programs. In contrast, domain-specific models such as \mono{}, which are pre-trained on MPI-related programming languages of C and C++, outperform larger models. 
Subsequently, we introduce a dedicated downstream task of MPI-based program generation by fine-tuning \mono{} on \mpicorpus{}. We call the resulting model as \mpir{}. We propose an innovative pre-processing for completion only after observing the whole code, thus enabling better completion with a wider context. Comparative analysis against GPT-3.5 zero-shot performance, using a novel HPC-oriented evaluation method, demonstrates that \mpir{} excels in generating accurate MPI functions calls. The success of this tailored solution underscores the importance of domain-specific fine-tuning in optimizing language models for parallel computing code generation, paving the way for a new generation of automatic parallelization tools.

The sources of this work are available at our GitHub \textcolor{blue}{\href{https://github.com/Scientific-Computing-Lab-NRCN/MPI-rigen}{\mpir{}}} repository.
\end{abstract}

\maketitle

\pagestyle{plain}

% \clearpage
\section{MPI: Source-to-Source Parallelization with Compilers}
\label{sec:previous_work}

Efforts in the domain of source-to-source automatic parallelization have primarily focused on transitioning from serial code to shared-memory parallelization, especially with OpenMP. Such approaches have relied on heuristics and rule-based methods (e.g., ComPar~\cite{mosseri2020compar}, Par4all~\cite{amini2012par4all, par4allhome}, and Cetus~\cite{bae2013cetus}) employing AST generation and data dependence algorithms. However, these methods often face challenges in handling diverse syntax and may yield sub-optimal results, as noted in several studies~\cite{harel2020source, milewicz2021negative, prema2017identifying, prema2019study}. 

Other marginal efforts have been made to transform shared memory to distributed memory parallelization, yet not directly from serial to distributed memory parallelization. OMP2MPI~\cite{OMP2MPI} is a source-to-source compiler based on the BSCs Mercurium \cite{balart2004nanos} framework that automatically generates MPI source code from shared-memory parallel OpenMP code. OMP2MPI gets parallel OpenMP code, and its AST as input, detects and transforms \texttt{pragma omp parallel for} blocks, and then divides the task into MPI manager and worker processes. CATO~\cite{CATO} is another static compiler that uses LLVM and Clang to transform OpenMP code to MPI. Its main component is an LLVM transform pass, which transforms the original OpenMP kernel using an intermediate representation (IR), an assembly-like representation of code.

These tools have many limitations. First, given the final purpose is to automatically convert serial code to a distributed one, converting serial code to a shared memory through the process leads to performance degradation due to imperfections in these compilers, including ones that originally related to the serial to shared memory automatic parallelization. Imperfections like increasing run time, using two-sided communication only, and even the necessity of external information ~\cite{hamidouche2011framework} damage the utility of this as a practical tool. Second, most of the static tools are source-to-source compilers, which obligate their use on compiled codes only.
\vspace{-0.3cm}

\section{MPI Parallelization with LLMs}

Recently, there has been a shift towards data-driven methods, especially large language models (LLMs). The basic approach to use LLMs for parallelization tasks is by utilizing those as-is in a zeroshot performance fashion~\cite{godoy2023evaluation, cuda, valero2023comparing, nichols2023modeling, mahmud2023autoparllm, nichols2024can}, while the more advanced approach is built upon fine-tuning said models for the parallelization specific task~\cite{harel2023learning, kadosh2023pragformer, kadosh2023advising, nichols2023modeling, chen2023lm4hpc, ding2023hpc, chen2023lm4hpc, chen2024ompgpt, kadosh2023scope}. In both of those cases, there has been success in utilizing transformers to \textit{classify} the need for OpenMP pragmas and MPI functions~\cite{schneider2023mpi} and even to \textit{generate} single-lined OpenMP pragmas~\cite{kadosh2023domain}, introducing a novel approach for automatic parallelization. However, the nuanced task of generating intricate, \textit{multi}-functional MPI codes across diverse \textit{locations} in a dedicated fine-tuned model has remained unexplored~\cite{chen2024position}. Nevertheless, recently, \mono{} \cite{kadosh2023domain} has been introduced as a novel approach aimed at enhancing LMs tailored for HPC tasks. The hypothesis posits that HPC-specific LMs, such as those designed and trained specifically on HPC datasets, would outperform existing LMs in HPC contexts. Two experiments are conducted to validate this hypothesis. Initially, \mono{} is constructed by reducing the number of layers of PolyCoder code LM by a factor of 4 and pretraining it solely on C and C++ codes. Despite its smaller size, \mono{} achieves comparable perplexity scores to PolyCoder on the \hpcorpus{} dataset, indicating its efficacy in understanding HPC-specific code structures. Subsequently, \mono{}'s performance on HPC tasks is evaluated, focusing on CodeBLEU competence for high-performance and parallel code generations, particularly in OpenMP parallelization tasks. Existing LMs struggle with capturing local semantics, affecting their performance in predicting OpenMP clauses accurately. To address this limitation, \tokom{}, a novel code pre-processing scheme that eliminates local semantics, is introduced. \mono{} consistently outperforms existing LMs across various tests and tasks, demonstrating its robustness and adaptability in HPC contexts, even under semantic-less settings. Yet, no benchmark or adaptation of \mono{} was performed for the MPI code generation problem.

\definecolor{mpiblue}{RGB}{0, 0, 255} % Define the color for MPI lines

\begin{figure*}[!h]
    \centering
    \begin{subfigure}[b]{0.3\textwidth}
\begin{lstlisting}[breaklines=true, breakatwhitespace=true, basicstyle=\footnotesize\ttfamily, columns=fullflexible]
int main(argc,argv)
{
    int done = 0, n = 10000, rank, numprocs, i;
    double pi_total, pi, h, sum, x;

    MPI_Init(&argc,&argv);
    MPI_Comm_size(MPI_COMM_WORLD, &numprocs);
    MPI_Comm_rank(MPI_COMM_WORLD, &rank);

    while (!done)
    {
	   MPI_Bcast(&n, 1, MPI_INT, 0, MPI_COMM_WORLD);
	   if (n == 0) 
              break;
	   h   = 1.0 / (double) n;...
\end{lstlisting}        
\caption{MPI Source code}
        \label{fig:subifg1}
    \end{subfigure}
        \begin{subfigure}[b]{0.33\textwidth}
    \begin{lstlisting}[breaklines=true, breakatwhitespace=true, basicstyle=\footnotesize\ttfamily, columns=fullflexible]
int func_270(type_255,type_762)
{
    int var_289 = num_463, var_649 = num_199, var_635, var_257, var_165;
    double var_663, var_792, var_305, var_610, var_55;

    MPI_Init(&type_255,&type_762);
    MPI_Comm_size(var_674,&var_257);
    MPI_Comm_rank(var_674,&var_635);

    while (!var_289)
    {
        MPI_Bcast(&var_649, num_700, var_167, num_463, var_674);
        if (var_649 == num_463) 
            break;
        var_305   = num_302 / (double) var_649;...
    \end{lstlisting}
        \caption{\tokom{} MPI source code}
        \label{fig:tokompiler}
    \end{subfigure}
    %\hspace{0.2cm}
    %\hfill
    %
    \begin{subfigure}[b]{0.3\textwidth}
    \begin{lstlisting}[breaklines=true, breakatwhitespace=true, basicstyle=\footnotesize\ttfamily, columns=fullflexible]
int main(argc,argv)
{
    int done = 0, n = 10000, rank, numprocs, i;
    double pi_total, pi, h, sum, x;

    while (!done)
    {
	   if (n == 0) 
              break;
	   h   = 1.0 / (double) n;...
    
    (6, MPI_Init(&argc,&argv);) |\Suppressnumber|
    (7, MPI_Comm_size(MPI_COMM_WORLD,&numprocs);)
    (8, MPI_Comm_rank(MPI_COMM_WORLD,&rank);) 
    (12, MPI_Bcast(&n, 1, MPI_INT, 0, MPI_COMM_WORLD);)...
    \end{lstlisting}
        \caption{MPI Pre-processed source code}
        \label{fig:subfig2}
    \end{subfigure}
    \caption{The MPI functions in the source code (a) are removed and concatenated with their corresponding line number to the last line (3). This way, \mpir{} learns in a left-to-right fashion the relation between code and its appropriate MPI functions. The \tokom{} version in (2), embeds AST information while neglecting any semantic information. \tokom{} version is used to demonstrate \mono{}'s semantic information independence during generation compared to other models.}
    \label{fig:preprocess}
\end{figure*}

\section{Research Objectives and Contributions}

The gap in the existing methods for MPI source-to-source code parallelization -- either by compilers or LMs -- has prompted our proposal for a data-driven generative language model using a sequence-to-sequence transformer-based approach. Such model's aim is to automatically suggest MPI functions for MPI codes with MPI functions excluded (semi-serial codes). This is by using a designated database of semi-serial codes and fully MPI parallelized code pairs, providing a new perspective on addressing the complexities of distributed-memory parallelization. Thus, this study presents \mpir{}, a tool developed to assist MPI programmers in automatically generating correct MPI functions in an MPI-based domain decomposition parallel code. \mono{} is the base model, hence, to check its suitability and \mpir{} performance we would like to answer the following research questions:
\begin{itemize}[leftmargin=*]
\item \textbf{RQ1:} Is \mono{} capable of generating proper MPI code without fine-tuning, and in the absence of local semantics?
\item \textbf{RQ2:} Is \mpir{} capable of inserting the calls to MPI functions in the right locations?
\item \textbf{RQ3:} Is \mpir{} capable of generating calls to correct MPI functions?
\item \textbf{RQ4:} Is \mpir{} capable of generating correct arguments to the MPI functions?
\end{itemize}

{\textbf{Contributions.}} The main contributions of this paper are:
\begin{itemize}[leftmargin=*]
    \item We create the first MPI codes only dataset --- \mpicorpus{} which is based on \mpicodecorpus{} and \hpcorpus{}.
    \item We have demonstrated {\mono} understands MPI codes better than PolyCoder and GPT3.5 by using a code completion task on \mpicorpus{} with \tokom{} source code version, a code with AST information embedded while semantic information is neglected.
    \item We propose an innovative pre-process for completion tasks merely after observing the whole code, thus enabling better completion with a wider context.
    \item We train and evaluate our approach, named {\mpir} by fine-tuning a domain-specific model --- {\mono}, and find that our model performs well in suggesting MPI functions for domain decomposition into an MPI-based parallel code and is better than GPT3.5 zero-shot.
\end{itemize}

\section{\mpicorpus{}}
While MPI programs have been already scraped from GitHub and gathered in corpora such as \mpicodecorpus{}~\cite{schneider2023mpi} or even \hpcorpus{}~\cite{kadosh2023quantifying}, no corpus contains merely MPI domain decomposition codes. \mpicorpus{} is a corpus consisting of both of the corpora above but under the filter of MPI domain decomposition codes only and with duplication removal resulting in a total of 16,384 programs.

% \begin{table}[htbp]
%     \begin{tabular}[t]{|c|c|}
%     \hline
%     \textbf{Function}  & \textbf{Amount}  \\
%     \hline
%      \texttt{MPI\_Finalize} & 19,183 \\ 
%      \texttt{MPI\_Init} & 16,135 \\
%      \texttt{MPI\_Comm\_rank} & 16,096 \\
%      \texttt{MPI\_Send} & 14,534 \\
%      \texttt{MPI\_Comm\_size}  & 14,387 \\
%      \texttt{MPI\_Recv}  & 13,783 \\
%      \texttt{MPI\_Bcast} & 6,995 \\
%      \texttt{MPI\_Reduce} & 3,600 \\
%     \hline
%     \end{tabular}
%     \caption{MPI Common Core functions distribution for \mpicorpus{} dataset.}
%     \label{table:functions_dist}
% \end{table}
\begin{table}[htbp]
    \centering
    \begin{tabular}[t]{|c|c|c|c|}
    \hline
    \textbf{Function}  & \textbf{Amount} & \textbf{Function}  & \textbf{Amount} \\
    \hline
     \texttt{MPI\_Finalize} & 19,183 & \texttt{MPI\_Comm\_size}  & 14,387 \\
     \texttt{MPI\_Init} & 16,135 & \texttt{MPI\_Recv}  & 13,783 \\
     \texttt{MPI\_Comm\_rank} & 16,096 & \texttt{MPI\_Bcast} & 6,995 \\
     \texttt{MPI\_Send} & 14,534 & \texttt{MPI\_Reduce} & 3,600 \\
    \hline
    \end{tabular}
    \caption{MPI Common Core functions distribution for \mpicorpus{} dataset.}
    \label{table:functions_dist}
\end{table}

\begin{figure}
    \centering
            \begin{tikzpicture}
            \begin{axis}[
                legend columns=2,
                ybar,
                bar width=0.15cm, % Adjust the width as needed
                ylabel={CodeBLEU},
                symbolic x coords={context-100, context-300, context-600},
                xtick=data,
                ymin=0,
                ymax=1.2,
                enlarge x limits=0.2,
                %legend pos=north east,
                %legend style={font=\scriptsize}
                nodes near coords,
                every node near coord/.append style={rotate=90, anchor=west, font=\scriptsize},
                ybar=3pt, 
                legend style={at={(0.5,-0.3)},anchor=north,legend columns=-1},
                grid style=dashed,
                height=4cm,
                width=\linewidth,
        ymajorgrids=true,
            ]
            \addplot[fill=blue!30] coordinates {
                (context-100, 0.563)
                (context-300, 0.684)
                (context-600, 0.815)
            };
            \addplot[fill=red!30] coordinates {
                (context-100, 0.496)
                (context-300, 0.692)
                (context-600, 0.785)
            };
            \addplot[fill=orange!30] coordinates {
                (context-100, 0.377)
                (context-300, 0.539)
                (context-600, 0.714)
            };
            \addplot[fill=white!30] coordinates {
                (context-100, 0.466)
                (context-300, 0.465)
                (context-600, 0.477)
            };
            \addplot[fill=yellow!30] coordinates {
                (context-100, 0.504)
                (context-300, 0.586)
                (context-600, 0.6)
            };
            \addplot[fill=green!30] coordinates {
                (context-100, 0.517)
                (context-300, 0.589)
                (context-600, 0.601)
            };

            \legend{\mono{}, \mono{}+Tokompiler, PolyCoder, PolyCoder+Tokompiler, GPT3.5, GPT3.5+Tokompiler}
            \end{axis}
        \end{tikzpicture}
    \caption{Performance of various models on Code Completion task over the \mpicorpus{}. Models predict code continuation starting from token 100, 300, and 600.}
    \label{fig:code_bleu}
\end{figure}
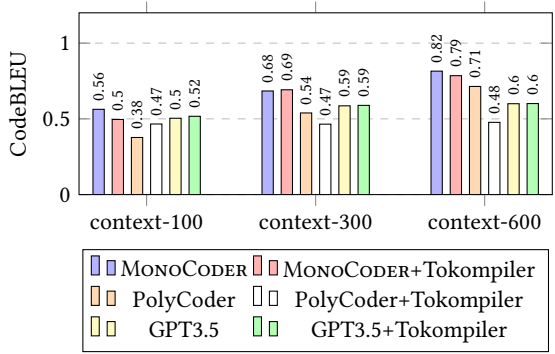
\begin{figure*}
\begin{subfigure}{0.33\textwidth}
    \centering
    \begin{tikzpicture}
    \begin{axis}[
        axis lines = left,
        ymin=0,
        ymax=1,
        bar width=0.15cm,
        xmin=1,
        xmax=20,
        ylabel={Accuracy},
        xlabel={\texttt{n}: Number of MPI functions calls},
        symbolic x coords={1, 2, 3, 4, 5, 6, 7, 8, 9, 10, 11, 12, 13, 14, 15, 16, 17, 18, 19, 20},
        xticklabels={1,,3,,5,,7,,9,,11,,13,,15,,17,,19,},
        xtick=data,
         width=\textwidth, % Adjusted width
        height=5cm,
        legend={},
        cycle list/Set2,
        every node near coord/.append style={opacity=1},
    ]
    \addplot+[line width=1pt] plot coordinates {(1,0.31) (2,0.44) (3,0.54) (4,0.55) (5,0.53) (6,0.53) (7,0.50) (8,0.50) (9,0.49) (10,0.48) (11,0.47) (12,0.47) (13,0.46) (14,0.46) (15,0.46) (16,0.45) (17,0.45) (18,0.45) (19,0.45) (20,0.45)};
    \addplot+[line width=1pt] plot coordinates {(1,0.47) (2,0.67) (3,0.72) (4,0.72) (5,0.71) (6,0.71) (7,0.68) (8,0.68) (9,0.67) (10,0.66) (11,0.65) (12,0.65) (13,0.64) (14,0.64) (15,0.64) (16,0.64) (17,0.63) (18,0.63) (19,0.63) (20,0.63)};
    \addplot+[line width=1pt] plot coordinates {(1,0.53) (2,0.75) (3,0.79) (4,0.80) (5,0.79) (6,0.78) (7,0.76) (8,0.75) (9,0.74) (10,0.74) (11,0.73) (12,0.72) (13,0.72) (14,0.71) (15,0.71) (16,0.71) (17,0.71) (18,0.71) (19,0.71) (20,0.71)};
    %\legend{Variance 0, Variance 1, Variance 2}
    \end{axis}
    \end{tikzpicture}
    \caption{Locations}
\end{subfigure}%
\hfill
\begin{subfigure}{0.33\textwidth}
    \centering
    \begin{tikzpicture}
    \begin{axis}[
        axis lines = left,
        legend columns=3,
        ymin=0.5,
        ymax=1,
        bar width=0.15cm,
        ylabel={Accuracy},
        xlabel={\texttt{n}: Number of MPI functions calls},
        symbolic x coords={1, 2, 3, 4, 5, 6, 7, 8, 9, 10, 11, 12, 13, 14, 15, 16, 17, 18, 19, 20},
        xticklabels={1,,3,,5,,7,,9,,11,,13,,15,,17,,19,},
        xtick=data,
         width=\textwidth, % Adjusted width
        height=5cm,
        legend style={at={(0.5,1.3)}, anchor=north, legend columns=-1},
        cycle list/Set2,
        every node near coord/.append style={opacity=1},
    ]
    \addplot+[line width=1pt] plot coordinates {(1,0.89) (2,0.57) (3,0.62) (4,0.65) (5,0.60) (6,0.61) (7,0.61) (8,0.62) (9,0.61) (10,0.61) (11,0.60) (12,0.60) (13,0.60) (14,0.60) (15,0.60) (16,0.60) (17,0.60) (18,0.60) (19,0.60) (20,0.60)};
    \addplot+[line width=1pt] plot coordinates {(1,0.83) (2,0.69) (3,0.73) (4,0.73) (5,0.70) (6,0.70) (7,0.68) (8,0.69) (9,0.69) (10,0.68) (11,0.67) (12,0.67) (13,0.66) (14,0.66) (15,0.66) (16,0.66) (17,0.66) (18,0.66) (19,0.66) (20,0.66)};
    \addplot+[line width=1pt] plot coordinates {(1,1.00) (2,0.78) (3,0.78) (4,0.77) (5,0.72) (6,0.72) (7,0.70) (8,0.71) (9,0.70) (10,0.69) (11,0.69) (12,0.68) (13,0.68) (14,0.67) (15,0.67) (16,0.67) (17,0.67) (18,0.67) (19,0.67) (20,0.67)};
    \legend{Variance 0, Variance 1, Variance 2}
    \end{axis}
    \end{tikzpicture}
    \caption{Functions}
\end{subfigure}%
\hfill
\begin{subfigure}{0.33\textwidth}
    \centering
    \begin{tikzpicture}
    \begin{axis}[
        axis lines = left,
        legend columns=1,
        ymin=0.8,
        ymax=1,
        bar width=0.1cm,
        ylabel={Accuracy},
        xlabel={\texttt{n}: Number of MPI functions calls},
        symbolic x coords={1, 2, 3, 4, 5, 6, 7, 8, 9, 10, 11, 12, 13, 14, 15, 16, 17, 18, 19, 20},
        xticklabels={1,,3,,5,,7,,9,,11,,13,,15,,17,,19,},
        xtick=data,
         width=\textwidth, % Adjusted width
        height=5cm,
        cycle list/Set2,
        every node near coord/.append style={opacity=1},
    ]
    \addplot+[line width=1pt] plot coordinates {(1,0.94) (2,0.96) (3,0.97) (4,0.98) (5,0.96) (6,0.96) (7,0.96) (8,0.96) (9,0.96) (10,0.95) (11,0.95) (12,0.95) (13,0.95) (14,0.95) (15,0.95) (16,0.95) (17,0.95) (18,0.95) (19,0.95) (20,0.95)};
    \addplot+[line width=1pt] plot coordinates {(1,0.94) (2,0.97) (3,0.98) (4,0.98) (5,0.96) (6,0.96) (7,0.95) (8,0.95) (9,0.95) (10,0.94) (11,0.94) (12,0.94) (13,0.94) (14,0.94) (15,0.94) (16,0.94) (17,0.94) (18,0.94) (19,0.94) (20,0.94)};
    \addplot+[line width=1pt] plot coordinates {(1,0.94) (2,0.97) (3,0.97) (4,0.97) (5,0.95) (6,0.95) (7,0.95) (8,0.95) (9,0.94) (10,0.94) (11,0.94) (12,0.94) (13,0.94) (14,0.94) (15,0.93) (16,0.93) (17,0.93) (18,0.93) (19,0.93) (20,0.93)};
    %\legend{Variance 0, Variance 1, Variance 2}
    \end{axis}
    \end{tikzpicture}
    \caption{Arguments}
\end{subfigure}%
\caption{Performance breakdown of \mpir{} (fine-tuned \mono{}-0.7B over 16K MPI codes from \mpicorpus{}) over programs containing (\texttt{n}) or less of MPI function calls (X axis). Y axis is accuracy obtained for such programs. Note shifted scales in sub-figure b and c.}
\label{fig:subfig1}
\end{figure*}
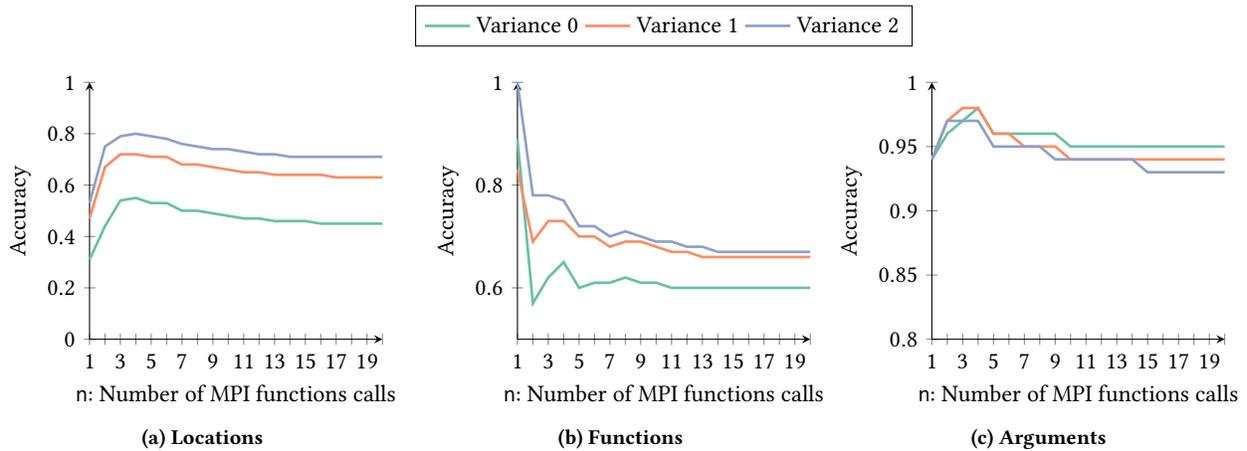

%%%%%%%%%%%%%%%%%%%%%%%%%%%%%%%%%%

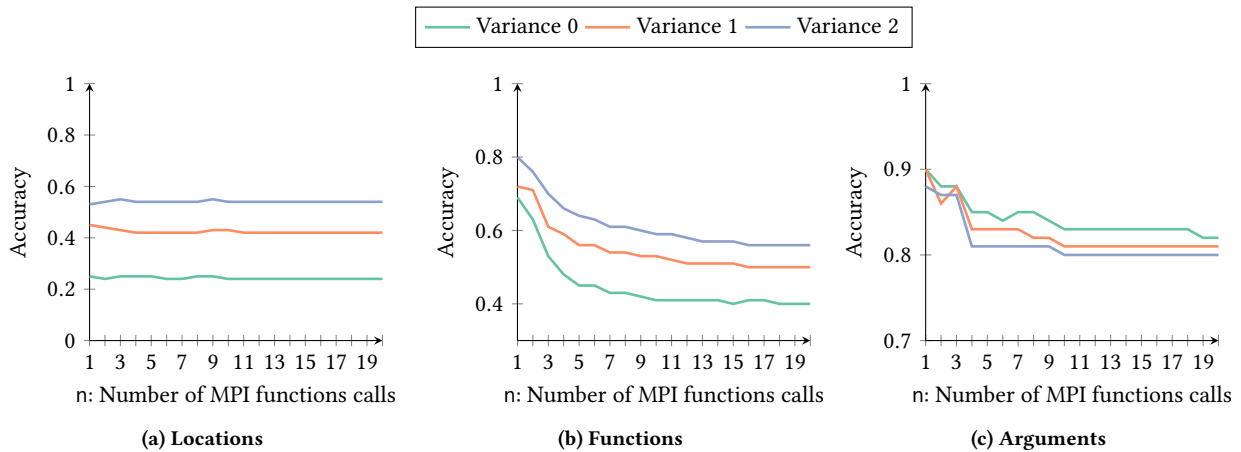
\begin{figure*}
\begin{subfigure}{0.33\textwidth}
    \centering
    \begin{tikzpicture}
    \begin{axis}[
        axis lines = left,
        legend columns=1,
        ymin=0,
        ymax=1,
        bar width=0.15cm,
        xmin=1,
        xmax=20,
        ylabel={Accuracy},
        xlabel={\texttt{n}: Number of MPI functions calls},
        symbolic x coords={1, 2, 3, 4, 5, 6, 7, 8, 9, 10, 11, 12, 13, 14, 15, 16, 17, 18, 19, 20},
        xticklabels={1,,3,,5,,7,,9,,11,,13,,15,,17,,19,},
        xtick=data,
         width=\textwidth, % Adjusted width
        height=5cm,
        cycle list/Set2,
        every node near coord/.append style={opacity=1},
    ]
    \addplot+[line width=1pt] plot coordinates {(1,0.25) (2,0.24) (3,0.25) (4,0.25) (5,0.25) (6,0.24) (7,0.24) (8,0.25) (9,0.25) (10,0.24) (11,0.24) (12,0.24) (13,0.24) (14,0.24) (15,0.24) (16,0.24) (17,0.24) (18,0.24) (19,0.24) (20,0.24)};
    \addplot+[line width=1pt] plot coordinates {(1,0.45) (2,0.44) (3,0.43) (4,0.42) (5,0.42) (6,0.42) (7,0.42) (8,0.42) (9,0.43) (10,0.43) (11,0.42) (12,0.42) (13,0.42) (14,0.42) (15,0.42) (16,0.42) (17,0.42) (18,0.42) (19,0.42) (20,0.42)};
    \addplot+[line width=1pt] plot coordinates {(1,0.53) (2,0.54) (3,0.55) (4,0.54) (5,0.54) (6,0.54) (7,0.54) (8,0.54) (9,0.55) (10,0.54) (11,0.54) (12,0.54) (13,0.54) (14,0.54) (15,0.54) (16,0.54) (17,0.54) (18,0.54) (19,0.54) (20,0.54)};
    %\legend{Variance 0, Variance 1, Variance 2}
    \end{axis}
    \end{tikzpicture}
    \caption{Locations}
\end{subfigure}%
\hfill
\begin{subfigure}{0.33\textwidth}
    \centering
    \begin{tikzpicture}
    \begin{axis}[
        axis lines = left,
        legend columns=3,
        ymin=0.3,
        ymax=1,
        bar width=0.15cm,
        ylabel={Accuracy},
        xlabel={\texttt{n}: Number of MPI functions calls},
        symbolic x coords={1, 2, 3, 4, 5, 6, 7, 8, 9, 10, 11, 12, 13, 14, 15, 16, 17, 18, 19, 20},
        xticklabels={1,,3,,5,,7,,9,,11,,13,,15,,17,,19,},
        xtick=data,
         width=\textwidth, % Adjusted width
        height=5cm,
        legend style={at={(0.5,1.3)}, anchor=north, legend columns=-1},
        cycle list/Set2,
        every node near coord/.append style={opacity=1},
    ]
    \addplot+[line width=1pt] plot coordinates {(1,0.69) (2,0.63) (3,0.53) (4,0.48) (5,0.45) (6,0.45) (7,0.43) (8,0.43) (9,0.42) (10,0.41) (11,0.41) (12,0.41) (13,0.41) (14,0.41) (15,0.40) (16,0.41) (17,0.41) (18,0.40) (19,0.40) (20,0.40)};
    \addplot+[line width=1pt] plot coordinates {(1,0.72) (2,0.71) (3,0.61) (4,0.59) (5,0.56) (6,0.56) (7,0.54) (8,0.54) (9,0.53) (10,0.53) (11,0.52) (12,0.51) (13,0.51) (14,0.51) (15,0.51) (16,0.50) (17,0.50) (18,0.50) (19,0.50) (20,0.50)};
    \addplot+[line width=1pt] plot coordinates {(1,0.80) (2,0.76) (3,0.70) (4,0.66) (5,0.64) (6,0.63) (7,0.61) (8,0.61) (9,0.60) (10,0.59) (11,0.59) (12,0.58) (13,0.57) (14,0.57) (15,0.57) (16,0.56) (17,0.56) (18,0.56) (19,0.56) (20,0.56)};
    \legend{Variance 0, Variance 1, Variance 2}
    \end{axis}
    \end{tikzpicture}
    \caption{Functions}
\end{subfigure}%
\hfill
\begin{subfigure}{0.33\textwidth}
    \centering
    \begin{tikzpicture}
    \begin{axis}[
        axis lines = left,
        legend columns=1,
        ymin=0.7,
        ymax=1,
        bar width=0.1cm,
        ylabel={Accuracy},
        xlabel={\texttt{n}: Number of MPI functions calls},
        symbolic x coords={1, 2, 3, 4, 5, 6, 7, 8, 9, 10, 11, 12, 13, 14, 15, 16, 17, 18, 19, 20},
        xticklabels={1,,3,,5,,7,,9,,11,,13,,15,,17,,19,},
        xtick=data,
         width=\textwidth, % Adjusted width
        height=5cm,
        cycle list/Set2,
        every node near coord/.append style={opacity=1},
    ]
    \addplot+[line width=1pt] plot coordinates {(1,0.90) (2,0.88) (3,0.88) (4,0.85) (5,0.85) (6,0.84) (7,0.85) (8,0.85) (9,0.84) (10,0.83) (11,0.83) (12,0.83) (13,0.83) (14,0.83) (15,0.83) (16,0.83) (17,0.83) (18,0.83) (19,0.82) (20,0.82)};
    \addplot+[line width=1pt] plot coordinates {(1,0.90) (2,0.86) (3,0.88) (4,0.83) (5,0.83) (6,0.83) (7,0.83) (8,0.82) (9,0.82) (10,0.81) (11,0.81) (12,0.81) (13,0.81) (14,0.81) (15,0.81) (16,0.81) (17,0.81) (18,0.81) (19,0.81) (20,0.81)};
    \addplot+[line width=1pt] plot coordinates {(1,0.88) (2,0.87) (3,0.87) (4,0.81) (5,0.81) (6,0.81) (7,0.81) (8,0.81) (9,0.81) (10,0.80) (11,0.80) (12,0.80) (13,0.80) (14,0.80) (15,0.80) (16,0.80) (17,0.80) (18,0.80) (19,0.80) (20,0.80)};
    %\legend{Variance 0, Variance 1, Variance 2}
    \end{axis}
    \end{tikzpicture}
    \caption{Arguments}
\end{subfigure}%
\caption{Performance breakdown of GPT3.5 using prompt ``Generate the optimal MPI functions for the provided code, and supply in the response the entire complete code with those MPI functions: [CODE]''. X axis represents programs containing \texttt{n} or less of MPI function calls. Y axis is accuracy obtained for such programs. Note shifted scales in sub-figure b and c.}
\label{fig:subfig3}

\end{figure*}

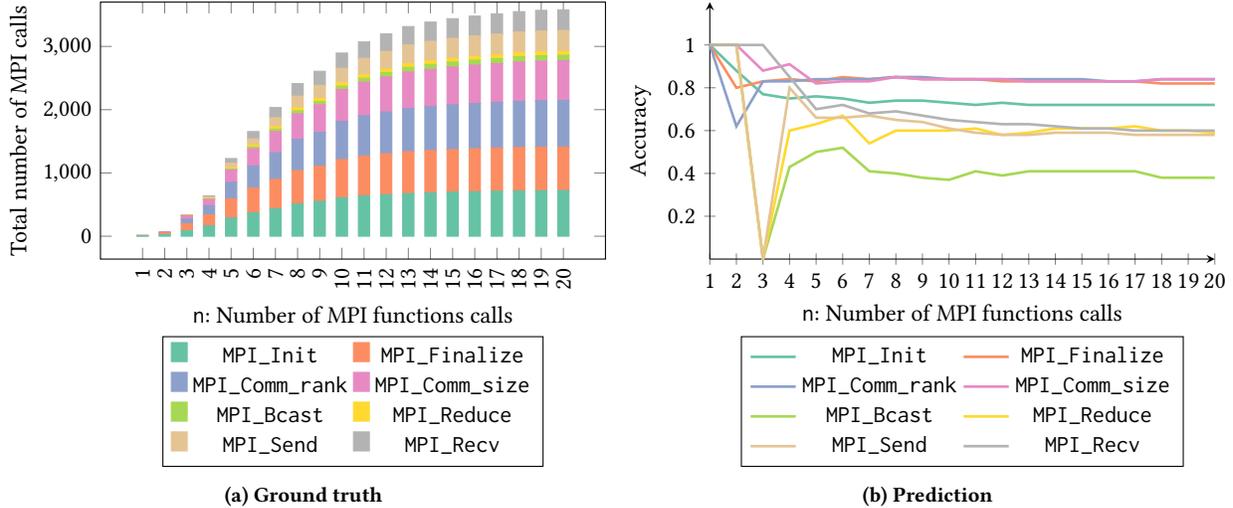
\begin{figure*}
    \begin{subfigure}{0.5\textwidth}
        \centering

\begin{tikzpicture}
    \begin{axis}[
        ybar stacked,
        legend columns=2,
        ymax=3700,
        bar width=0.15cm,
        ylabel={Total number of MPI calls},
        xlabel={\texttt{n}: Number of MPI functions calls},
        symbolic x coords={1, 2, 3, 4, 5, 6, 7, 8, 9, 10, 11, 12, 13, 14, 15, 16, 17, 18, 19, 20},
        xtick=data,
        xticklabel style={rotate=90},
         width=\textwidth, % Adjusted width
        height=5cm,
        legend style={at={(0.5,-0.3)}, anchor=north, legend columns=-1},
        cycle list/Set2,
        every node near coord/.append style={opacity=1},
    ]

     \addplot+[ybar,fill] plot coordinates {(1,8) (2,25) (3,90) (4,163) (5,287) (6,367) (7,437) (8,509) (9,551) (10,608) (11,637) (12,658) (13,675) (14,687) (15,697) (16,704) (17,711) (18,716) (19,721) (20,723)};
    \addplot+[ybar,fill] plot coordinates {(1,9) (2,30) (3,108) (4,176) (5,300) (6,395) (7,461) (8,529) (9,556) (10,604) (11,631) (12,647) (13,662) (14,668) (15,672) (16,679) (17,684) (18,687) (19,687) (20,687)};
    \addplot+[ybar,fill] plot coordinates {(1,0) (2,8) (3,72) (4,143) (5,264) (6,349) (7,422) (8,496) (9,533) (10,605) (11,637) (12,659) (13,681) (14,693) (15,707) (16,715) (17,721) (18,729) (19,734) (20,737)};
    \addplot+[ybar,fill] plot coordinates {(1,0) (2,1) (3,56) (4,99) (5,197) (6,271) (7,336) (8,401) (9,437) (10,502) (11,530) (12,553) (13,578) (14,589) (15,599) (16,607) (17,613) (18,621) (19,626) (20,629)};
    %MPI_Bcast
    \addplot+[ybar,fill] plot coordinates {(1,0) (2,0) (3,1) (4,7) (5,14) (6,27) (7,37) (8,47) (9,52) (10,54) (11,63) (12,69) (13,74) (14,78) (15,78) (16,79) (17,79) (18,85) (19,85) (20,85)};
    %MPI_Reduce
    \addplot+[ybar,fill] plot coordinates {(1,0) (2,0) (3,1) (4,10) (5,19) (6,27) (7,37) (8,47) (9,50) (10,50) (11,51) (12,53) (13,54) (14,56) (15,56) (16,57) (17,58) (18,60) (19,60) (20,61)};
    %MPI_Send
    \addplot+[ybar,fill] plot coordinates {(1,0) (2,0) (3,1) (4,20) (5,73) (6,104) (7,144) (8,186) (9,207) (10,232) (11,263) (12,281) (13,299) (14,311) (15,319) (16,323) (17,328) (18,331) (19,331) (20,331)};
    %MPI_Recv
    \addplot+[ybar,fill] plot coordinates {(1,0) (2,0) (3,1) (4,20) (5,74) (6,114) (7,160) (8,197) (9,221) (10,240) (11,259) (12,278) (13,291) (14,302) (15,311) (16,315) (17,321) (18,321) (19,326) (20,326)};
    
    \legend{\texttt{MPI\_Init}, \texttt{MPI\_Finalize}, \texttt{MPI\_Comm\_rank}, \texttt{MPI\_Comm\_size}, \texttt{MPI\_Bcast}, \texttt{MPI\_Reduce}, \texttt{MPI\_Send}, \texttt{MPI\_Recv}}
    \end{axis}
\end{tikzpicture}
    \caption{Ground truth}
    \label{fig:analysis_ground_truth}
    \end{subfigure}%
    \begin{subfigure}{0.5\textwidth}
        \centering
    \begin{tikzpicture}
    \begin{axis}[
        axis lines = left,
        legend columns=2,
        ymin=0,
        ymax=1.2,
        ytick={0.2,0.4,0.6,0.8,1.0},
        ylabel={Accuracy},
        xlabel={\texttt{n}: Number of MPI functions calls},
        %symbolic x coords={1, 2, 3, 4, 5, 6, 7, 8, 9, 10, 11, 12, 13, 14, 15, 16, 17, 18, 19, 20},
        %symbolic x coords={0.25, 0.5, 0.75, 1},
        xtick=data,
         width=\textwidth, % Adjusted width
        height=5cm,
        legend style={at={(0.5,-0.3)}, anchor=north, legend columns=-1},
        cycle list/Set2,
        every node near coord/.append style={opacity=1},
    ]

    \addplot+[line width=1pt] plot coordinates {(1,1.00) (2,0.88) (3,0.77) (4,0.75) (5,0.76) (6,0.75) (7,0.73) (8,0.74) (9,0.74) (10,0.73) (11,0.72) (12,0.73) (13,0.72) (14,0.72) (15,0.72) (16,0.72) (17,0.72) (18,0.72) (19,0.72) (20,0.72)};
    \addplot+[line width=1pt] plot coordinates {(1,1.00) (2,0.80) (3,0.83) (4,0.84) (5,0.83) (6,0.85) (7,0.84) (8,0.85) (9,0.84) (10,0.84) (11,0.84) (12,0.83) (13,0.83) (14,0.83) (15,0.83) (16,0.83) (17,0.83) (18,0.82) (19,0.82) (20,0.82)};
    \addplot+[line width=1pt] plot coordinates {(1,1.00) (2,0.62) (3,0.83) (4,0.83) (5,0.84) (6,0.84) (7,0.84) (8,0.85) (9,0.85) (10,0.84) (11,0.84) (12,0.84) (13,0.84) (14,0.84) (15,0.84) (16,0.83) (17,0.83) (18,0.84) (19,0.84) (20,0.84)};
    \addplot+[line width=1pt] plot coordinates {(1,1.00) (2,1.00) (3,0.88) (4,0.91) (5,0.82) (6,0.83) (7,0.83) (8,0.85) (9,0.84) (10,0.84) (11,0.84) (12,0.84) (13,0.83) (14,0.83) (15,0.83) (16,0.83) (17,0.83) (18,0.84) (19,0.84) (20,0.84)};
    %MPI_Bcast
    \addplot+[line width=1pt] plot coordinates {(1,1.00) (2,1.00) (3,0.00) (4,0.43) (5,0.50) (6,0.52) (7,0.41) (8,0.40) (9,0.38) (10,0.37) (11,0.41) (12,0.39) (13,0.41) (14,0.41) (15,0.41) (16,0.41) (17,0.41) (18,0.38) (19,0.38) (20,0.38)};
    %MPI_Reduce
    \addplot+[line width=1pt] plot coordinates {(1,1.00) (2,1.00) (3,0.00) (4,0.60) (5,0.63) (6,0.67) (7,0.54) (8,0.60) (9,0.60) (10,0.60) (11,0.61) (12,0.58) (13,0.59) (14,0.61) (15,0.61) (16,0.61) (17,0.62) (18,0.60) (19,0.60) (20,0.59)};
    %MPI_Send
    \addplot+[line width=1pt] plot coordinates {(1,1.00) (2,1.00) (3,0.00) (4,0.80) (5,0.66) (6,0.66) (7,0.67) (8,0.65) (9,0.64) (10,0.61) (11,0.59) (12,0.58) (13,0.58) (14,0.59) (15,0.59) (16,0.59) (17,0.58) (18,0.58) (19,0.58) (20,0.58)};
    %MPI_Recv
    \addplot+[line width=1pt] plot coordinates {(1,1.00) (2,1.00) (3,1.00) (4,0.85) (5,0.70) (6,0.72) (7,0.68) (8,0.69) (9,0.67) (10,0.65) (11,0.64) (12,0.63) (13,0.63) (14,0.62) (15,0.61) (16,0.61) (17,0.60) (18,0.60) (19,0.60) (20,0.60)};
    
    \legend{\texttt{MPI\_Init}, \texttt{MPI\_Finalize}, \texttt{MPI\_Comm\_rank}, \texttt{MPI\_Comm\_size}, \texttt{MPI\_Bcast}, \texttt{MPI\_Reduce}, \texttt{MPI\_Send}, \texttt{MPI\_Recv}}
    \end{axis}
\end{tikzpicture}
\caption{Prediction}
\label{fig:analysis_pred}
    \end{subfigure}
    \caption{Stacked bar chart of the ground truth and \mpir{} prediction distribution of selected MPI functions under variance 2 (correct location and function predictions are presented). X axis represents programs containing \texttt{n} or less of MPI function calls.}
    \label{fig:hist}
\end{figure*}

%\subsection{Dataset}
To create the dataset for training \mpir{} out of the corpus, several pre-process stages have been done (\autoref{fig:preprocess}):
\begin{enumerate}[leftmargin=*]
    \item Codes with both \texttt{MPI\_Init} and \texttt{MPI\_Finalize} and under the limit of 2048 tokens (with BPE tokenizer) have been added.
    \item Every line of each code snippet has been numbered.
    \item MPI functions through the code have been removed.
    \item Locations of the removed MPI functions and the functions themselves have been written to the last line of the codes.
\end{enumerate}
For convenience, the final dataset (with 13,322 programs) contains three fields -- ``Program'', ``Code'', ``MPI label'' -- and corresponds to the program's GitHub username, the code with removed MPI functions, and the label of the locations and their MPI functions.

``MPI Common Core'' functions, as defined in~\cite{schneider2023mpi}, are the most prevalent MPI functions in domain decomposition. Therefore, it is essential to validate that the distribution in the dataset is reasonable both for model training and evaluation purposes. Towards that end, we analyzed \mpicorpus{} for MPI Common Core functions. The distribution of these functions is presented in ~\autoref{table:functions_dist}.

\section{\mpir{} -- Experimental Results}
\label{sec:mpirigen}
This study presents \mpir{}, a tool developed to assist MPI programmers in automatically generating correct MPI functions in an MPI-based domain decomposition parallel code.
We introduce a dedicated downstream task: fine-tuning \mono{}, which is a PolyCoder model pre-trained for the C and C++ languages associated with MPI, on \mpicorpus{}, resulting in the creation of \mpir{}. 

Furthermore, we propose an innovative pre-process for the given completion task and completion tasks in general. A number specifying its line number is added for each line in the code. Then, all the MPI functions with their locations are removed and concatenated to the last line (\autoref{fig:preprocess}). Training input will be the resulting code and the test will be to predict the last line. This enables training a language model in a left-to-right fashion, outputting the MPI functions merely after observing the whole code, thus enabling better completion with a wider context. \mono{} is fine-tuned this way resulting in \mpir{}.

we ran two experiments: the first experiments evaluates ability of base models to generate MPI functions and answers \textbf{RQ1}, while the second experiment is designed to answer \textbf{RQ2}, \textbf{RQ3}, \textbf{RQ4} and it is the novel approach we are introducing.

\textbf{\textit{RQ1: MPI code generation.}} To develop \mpir{} and answer the first RQ, we first investigated the performance of state-of-the-art pre-trained code language models (PolyCoder and GPT3.5) in generating MPI codes using varied context sizes for next-token predictions compared to \mono{}. This code completion task has been done over a thousand examples with up to 2048 tokens in \mpicorpus{} and under an initial context of 100, 300, and 600 tokens (\autoref{fig:code_bleu}). In addition, to check the resilience of these pre-trained models to semantic information, the same test has been conducted through codes' \tokom{} \cite{kadosh2023domain} version. Any semantic information has been replaced using \tokom{} (\autoref{fig:tokompiler}) and thus, a measure of reliance of these pre-trained models on semantic information has been tested. Results show that as context grows, \mono{} is significantly outperforming PolyCoder and GPT-3.5 results, while also proving to not fix on local semantics for those results, suggesting high generality capabilities.

\textbf{\textit{Evaluation metric for RQ2, RQ3, RQ4.}} Since we evaluate MPI function generation, prevalent generation metrics such as BLUE, Meteor, and Rouge-l are not relevant. Thus, we propose an HPC-oriented evaluation method. First, functions matching locations are to be found. Matching toleration is determined by a variance. The variance is a flexibility measure for accuracy, as many times, the right functions appear in one or two lines from the original locations, usually without interfering with the original code structure (variance zero refers to exact location predictions). MPI functions with correct locations will be forwarded to calling evaluation, checking whether the right MPI function has been called. Afterward, arguments will be checked out of the correct MPI functions. Scoring is made by calculating the correct to total arguments ratio. This evaluation was conducted with a variance of 0-2 and as a number of MPI functions which reveals the accuracy distribution through simple MPI codes to more complicated ones. Functions accuracy is measured for matched locations only and arguments accuracy for matched functions only. The results are given as a function of the number of original MPI calls (1-20), showcasing that accuracy drops as complexity grows (besides the ultra simplicity of one or two MPI function calls).

\textbf{\textit{Results for RQ2, RQ3, RQ4.}} We developed the model using the Huggingface framework and trained it on two NVidia Tesla-V100 GPUs, each having 32GB memory. The training was carried out with a batch size of 2, 2048 tokens and 3 epochs. As shown in \autoref{fig:subfig1}, for \mpir{}, the location as well as functions accuracy under variance of 2 reach 80\% and converge to 70\%, while for arguments it reaches 95\% and converge to 94\%. For GPT3.5 on the contrary, the accuracy converged to 50\% for locations and approximately 55\% and 80\% for functions and arguments resp (\autoref{fig:subfig3}).

\textbf{\textit{Result analysis and future work.}} We analyzed the performance of \mpir{} in details by breaking down the location and function accuracy over commonly-used MPI functions.
This analysis is presented in \autoref{fig:hist}, with \autoref{fig:analysis_ground_truth} showing the number of calls of different MPI functions in the ground truth and \autoref{fig:analysis_pred} showing the accuracy of predicting those calls with a variance of two. As the figure shows the accuracy of predicting the correct location varies for different MPI functions, and we believe it correlates with the proportion of different MPI calls in the training dataset and possibly in open-source repositories on GitHub. It would be thus an interesting study to develop a balanced MPI dataset and train \mpir{} on it. Currently, we do not evaluate the correctness of generated MPI codes, nonetheless we are envisioning an approach based on compilation and output verification as a part of immediate future work.

%High accuracy could be a misleading metric since the model might excel in merely one frequent function, \texttt{MPI\_Init}, for example. Therefore, accuracy alone can not guarantee a generalized operating model and matching function distribution is necessary.

\section{Conclusion}
Our findings reveal that widely used models like GPT-3.5 and specialized multi-lingual code models like PolyCoder exhibit notable performance degradation when generating MPI codes compared to their outcomes for general-purpose codes, as shown in \cite{kadosh2023domain}. In contrast, domain-specific models like \mono{}, pre-trained for the C and C++ languages associated with MPI, outperform larger models, showcasing high generality capabilities, especially when local misleading semantics are mitigated. Comparative analysis of \mpir{} against GPT zero-shot performance, using the above evaluation method for MPI functions generation, demonstrates that \mpir{} excels in generating accurate MPI codes. The success of this tailored solution underscores the importance of domain-specific fine-tuning in optimizing language models for parallel computing code generation.  

%\clearpage
\begin{acks}
This research was supported by the Israeli Council for Higher Education (CHE) via the Data Science Research Center, Ben-Gurion University of the Negev, Israel; Intel Corporation (oneAPI CoE program); and the Lynn and William Frankel Center for Computer Science. Computational support was provided by the NegevHPC project \cite{negevhpc} and Intel Developer Cloud~\cite{intel-cloud}. The authors thank Israel Hen and Gabi Dadush for their help and support.
\end{acks}
\bibliographystyle{ACM-Reference-Format}
\bibliography{sample-base}
\end{document}